
\input phyzzx.tex
\hoffset=1cm
\overfullrule0pt
\hsize=35.5pc \vsize=51pc
\normalparskip=0pt
\parindent=36pt
\itemsize=36pt
\def\ifmath#1{\relax\ifmmode #1\else $#1$\fi}

\def\hl{h^0}
\def\mhl{m_{\hl}}

\def\rta{\rightarrow}
\def\bold#1{\setbox0=\hbox{$#1$}%
     \kern-.025em\copy0\kern-\wd0
     \kern.05em\copy0\kern-\wd0
     \kern-.025em\raise.0433em\box0 }

\def\GENITEM#1;#2{\par\vskip6pt \hangafter=0 \hangindent=#1
   \Textindent{$ #2$ }\ignorespaces}

\def\unlock{\catcode`@=11} 
\def\lock{\catcode`@=12} 
\unlock
\def\ls#1{_{\lower1.5pt\hbox{$\scriptstyle #1$}}}
\def\chapter#1{\par \penalty-300 \vskip\chapterskip
   \spacecheck\chapterminspace
   \chapterreset \leftline{\bf \chapterlabel.~~#1}
   \nobreak\vskip\headskip \penalty 30000
   {\pr@tect\wlog{\string\chapter\space \chapterlabel}} }
\def\section#1{\par \ifnum\lastpenalty=30000\else
   \penalty-200\vskip\sectionskip \spacecheck\sectionminspace\fi
   \gl@bal\advance\sectionnumber by 1
   {\pr@tect
   \xdef\sectionlabel{\chapterlabel.%
       \the\sectionstyle{\the\sectionnumber}}%
   \wlog{\string\section\space \sectionlabel}}%
   \noindent {\it\sectionlabel.~~#1}\par
   \nobreak\vskip\headskip \penalty 30000 }\lock
\def\CERN{\centerline {\it CERN, TH-Division}
  \centerline{\it CH-1211 Geneva 23, Switzerland}}
\Pubnum={CERN-TH/95-178 \cr SCIPP 95/31}
\date={June, 1995}
\titlepage
\singlespace
\vbox to 1cm{}
\centerline{{\fourteenbf IS {$\bold{M_t\simeq M_W}$} RULED OUT?}
\foot{Work supported in part by the U.S.~Department of Energy.}}
\vskip1cm
\centerline{{\caps Howard E. Haber}\foot{Permanent address:
Santa Cruz Institute for Particle Physics, University of California,
Santa Cruz, CA 94064  USA.}}
\vskip .1in
\CERN
\vskip1cm
\vbox{ \narrower
\centerline{\bf Abstract}
\vskip6pt
A four generation supersymmetric model is proposed, in which the
Tevatron ``top-quark'' events are reinterpreted as the
production of $t^\prime$
which decays dominantly to $bW^+$.  In this model, $m_t\simeq m_W$,
and $t\rta\widetilde t\widetilde\chi^0_1$, with $\widetilde t\rta
c\widetilde\chi^0_1$.  This decay chain, which
rarely produces a hard isolated
lepton, would have been missed in all previous top quark searches.
A narrow region of the model parameter space exists
which cannot yet be ruled out by present data.  This model predicts
a rich spectrum of new physics which can be probed at LEP-II and the
Tevatron.
}
\vfill
\centerline{Invited talk presented at the XXX{\it th} Rencontres de Moriond,}
\centerline{``Electroweak Interactions and Unified Theories'',}
\centerline{Les Arcs, Savoie, France, 11--18 March, 1995.}
\vfill
\endpage

\chapter{Introduction}
\medskip

\REF\topquark{F. Abe \etal\ [CDF Collaboration], {\sl Phys. Rev. Lett.}
{\bf 74} (1995) 2626;
S. Abachi \etal\ [D0 Collaboration], {\sl Phys. Rev. Lett.}
{\bf 74} (1995) 2632.}
\REF\lepglobal{M.G. Alviggi \etal\ [LEP Electroweak Working Group],
LEPEWWG/95-01 (1995); M. Calvi, invited talk at this meeting.}
\REF\rhoparm{M. Veltman, {\sl Nucl. Phys.} {\bf B123} (1977) 89.}
\REF\chw{M. Carena, H.E. Haber and C.E.M. Wagner, CERN preprint in
preparation.}
Recently, the CDF and D0 Collaborations have announced the discovery
of the top quark at the Tevatron,\refmark\topquark\ with a measured
mass of $m_t=176\pm 8\pm 10$~GeV and
$m_t=199^{+19}_{-21}\pm 22$~GeV, respectively.  Both measurements
are in excellent agreement with the top quark mass deduced by the LEP
global analysis of precision electroweak
measurements.\refmark\lepglobal\  The LEP determination of $m_t$ is based
on the sensitivity of electroweak observables in $Z$ decay
to virtual top quark exchange, which
enters in two distinct ways.  First, top quark loops
in gauge boson self-energies (the so-called oblique corrections)
can directly effect the properties of the $Z$.  The most
famous of the oblique corrections is the top-quark contribution to
the electroweak $\rho$ parameter,\refmark\rhoparm\ which is given by
$\rho=1+\delta\rho$, where
$\delta\rho\simeq 3G_F m_t^2/8\pi^2\sqrt{2}$.
Second, virtual top quark exchange can
contribute to certain vertex radiative corrections.
For example, the one-loop correction to $Z\rta b\bar b$
is also quadratically sensitive to the top quark mass.
The LEP global fit yields $m_t=176\pm 10^{+17}_{-19}$~GeV,
where the second set of
errors corresponds to varying the Higgs mass between 60~GeV and 1~TeV
(with a central value of 300~GeV).

Clearly a heavy top quark mass has been confirmed.
But is there an alternative interpretation?
In this paper, I present a model constructed in
collaboration with Marcela Carena and Carlos Wagner,\refmark\chw\
in which we
explore the possibility of circumventing the apparent
ironclad conclusion that $m_t\gg m_W$.

\chapter{A Four Generation Supersymmetric Model with a Light Top Quark}
\medskip

Consider that the LEP measured rate for
$Z\rta b\bar b$ differs from the Standard Model prediction by
2.4$\sigma$.  Defining $R_b\equiv\Gamma(Z\rta
b\bar b)/\Gamma(Z\rta{\rm hadrons})$,\refmark\lepglobal\
$$R_b=\cases{0.2204\pm 0.0020,&LEP global fit;\cr
0.2157,&Standard Model prediction.\cr}
\eqn\zbbnumbers$$
Clearly, one does not give up on the Standard Model because of a
2.4$\sigma$ discrepancy.  Nevertheless, it is amusing to note that if
one extracts the top quark mass from this
measurement alone, one would conclude that $m_t<m_W$!  We proceed by
fixing $m_t\simeq m_W$ in what follows.  Of course, with such a light
top quark mass, we must address three obvious questions:
\pointbegin
Would not a top quark with $m_t\simeq m_W$ have already been
discovered at hadron colliders?
\point
What is the particle recently announced by CDF and D0 which is
observed to decay into $bW$?
\point
What is the nature of the new physics that contributes to the oblique
corrections and simulates the heavy top quark inferred by the LEP
experiments?

\REF\wwidth{B. Klima, invited talk at this meeting.}
\REF\dzerotop{S. Abachi \etal\ [D0 Collaboration], {\sl Phys. Rev. Lett.}
{\bf 72} (1994) 2138.}
\noindent
If the top quark were
sufficiently light, then $W^+\rta t\bar b$ would be kinematically
allowed; this would modify the total width of the $W$.  But
$\Gamma_W$ can be measured at hadron colliders indirectly by
studying the ratio of production cross section times
leptonic branching ratio of the
$W$ and $Z$.  The most recent analysis of this kind, reported at this
meeting by the D0 collaboration,\refmark\wwidth\ finds $m_t>62$~GeV.
Direct searches for the top quark at hadron colliders assume that an
observable fraction of top quark decays results in a final state
lepton.  For example, in ref.~\dzerotop,
the D0 collaboration ruled out the mass range
$m_W+m_b\lsim m_t<131$~GeV, assuming that the decay $t\rta bW$ is
not unexpectedly suppressed.\refmark\dzerotop\
Previous top quark searches at hadron
colliders are able to close the window between 62 and 85~GeV,
assuming that $t\rta bW^\star$ is the dominant top-quark decay mode.
However, in this case the final state is three-body since $W^\star$
is virtual.  If the top quark were to possess any two-body decay modes (due
to new physics processes), and if these modes rarely produced
leptons, then a top quark in this mass region would not have been
detected in any experiment.

\REF\pdg{Limits on supersymmetric particle masses
are summarized in L. Montanet \etal\ [Particle Data Group],
{\sl Phys. Rev.} {\bf D50} (1994) 1173.}
\REF\stopsearch{A. White, invited talk given at the SUSY-95 Conference,
15--19 May 1995, Palaiseau, France.}
An example of such a scenario occurs in supersymmetric models
in which the decay
$t\rta\widetilde t\widetilde\chi^0_1$ is kinematically allowed
(where $\widetilde t$ is the top squark and
$\widetilde\chi^0_1$ is the lightest neutralino).  Experimental searches for
both $\widetilde t$ and $\widetilde\chi^0_1$ place constraints on their
masses, but do not rule out the possibility of $M_{\tilde t}+
M_{\tilde\chi_1^0}<m_W$.
In particular, the LEP neutralino and chargino searches\refmark\pdg\
obtain a limit on the lightest neutralino mass which typically lies
between 20 and 25 GeV.  Using this result and the limits on the top squark
mass from searches at LEP and at the Tevatron,\refmark\stopsearch\
one finds that the mass region
$42\lsim M_{\tilde t}\lsim 60$~GeV cannot be excluded.

\REF\rudaz{I.I. Bigi and S. Rudaz, {\sl Phys. Lett.} {\bf 153B} (1985)
335.}
To be definite, we choose $m_t\simeq m_W$, $M_{\tilde t}\simeq 50$~GeV and
$M_{\tilde\chi^0_1}\simeq 25$~GeV.  Then, the dominant decay chain is
$t\rta\widetilde t\widetilde\chi^0_1$ followed by
$\widetilde t\rta c\widetilde\chi^0_1$ through a one-loop
process,\refmark\rudaz\  which rarely produces a hard isolated lepton.
Hence, these events would not have been detected at hadron colliders.
But, now we must
reconsider to the recent CDF and D0 discoveries and the LEP
``measurement'' of $m_t$.  We propose to account for these results
by introducing a fourth generation of quarks (and leptons) plus
their supersymmetric partners.  Then, $t^\prime\rta bW^+$ can be the
source of the CDF and D0 events, while the effects of
the third and fourth generation quarks and squarks contributing to
the oblique corrections are large enough
to be consistent with LEP precision electroweak
data.

\chapter{Phenomenological and Theoretical Constraints}
\medskip

The model parameters are determined by imposing the phenomenological
and theoretical constraints listed below.

1. In order that the $t^\prime$ be consistent with the CDF and
D0 ``top-quark'' events, its dominant decay must be $t^\prime\rta
bW^+$.  This means that $t^\prime\rta b^\prime W^\star$ must be a
three-body decay.  Furthermore, the $t^\prime$--$b$ mixing angle
($V_{t^\prime b}$) must not be too small; otherwise, the latter decay
will dominate.  We find:
$${\Gamma(t^\prime\rta b^\prime W^\star)\over\Gamma(t^\prime\rta b^\prime W)}
={9G_Fm_{t^\prime}^2\over \pi^2\sqrt{2}|V_{t^\prime b}|^2}
\int_0^{1-2\sqrt{x}+x}\,{z(1-z+x)\sqrt{(1-z+x)^2-4x}\over
\left(1-{zm_{t^\prime}^2/m_W^2}\right)^2}\,dz\,,
\eqn\gammarats$$
where $x\equiv m_{b^\prime}^2/m_{t^\prime}^2$.
Since the rate of the CDF and D0 ``top-quark'' events is consistent
with the QCD prediction for $t\bar t$ production under the assumption
that $BR(t\rta bW^+)=$ 100\%, a reinterpretation of these events as
$t^\prime \bar {t^\prime}$ production (followed by $t^\prime\rta bW^+$)
requires $BR(t^\prime\rta bW^+)$ to be near 1.  We assume that $V_{t^\prime b}$
lies between $V_{cb}=0.04$ and $V_{ud}=0.2$;  for definiteness, we
choose $V_{t^\prime b}=0.1$.  Then, if we require $BR(t^\prime\rta bW^+)
\gsim 0.75$, it follows that we must take $m_{b^\prime}\geq 105$~GeV.

\REF\irfps{C.T. Hill, {\sl Phys. Rev.} {\bf D24} (1981) 691;
C.T. Hill, C.N. Leung and S. Rao, {\sl Nucl. Phys.} {\bf B262} (1985)
517.}
2. In low-energy
supersymmetric model building, it is common practice to require
that all couplings of the model stay perturbative up to very high energies.
Here, we shall insist that the Higgs-quark Yukawa couplings
do not blow up below the grand unification (GUT) scale.  Then, if we wish to
have the $t^\prime$ and $b^\prime$ masses as large as possible, it
follows that the corresponding Yukawa couplings will be forced to lie close to
their quasi-infrared fixed points.\refmark\irfps\
For example, if we take $m_{t^\prime}
\geq 170$~GeV, then we find that $m_{b^\prime}\leq 110$~GeV.
Combined with point 1, we see that the mass of the $b^\prime$ is
essentially fixed.  Moreover, since we are at the infrared fixed point
values of the Yukawa couplings, which depend on the corresponding
masses and the ratio of Higgs vacuum expectation values, $\tan\beta$,
it follows that $\tan\beta$ is also fixed.  In this work, we choose
$m_{t^\prime}=170$~GeV and $m_{b^\prime}=110$~GeV; for these values
$\tan\beta\simeq 1.6$.  One can also add in the requirement that the
fourth generation leptons lie at their quasi-infrared fixed points
(in order to maximize their masses). We assume that the fourth
generation neutrino ($N$) is a Dirac fermion.  Then, the resulting lepton
masses are: $m_{\tau^\prime}\simeq 50$~GeV and $m_N
\simeq 80$~GeV.  Remarkably, these masses lie above the corresponding
bounds from LEP.  In addition, it is amusing to note that the above
masses are consistent with the unification of
all {\it four} fermion-Higgs Yukawa couplings at the GUT scale!
\endpage

\REF\poketal{M. Carena, M. Olechowski, S. Pokorski, and C.E.M. Wagner,
{\sl Nucl. Phys.} {\bf B419} (1994) 213; {\bf B426} (1994) 269.}
3. In order that $M_{\tilde t}<m_t$, there must be substantial
$\widetilde t_L$--$\widetilde t_R$ mixing.  The squared
mass of $\widetilde t$ is given by the smallest eigenvalue of the
matrix
$$\pmatrix{M_{\widetilde Q}^2+m_t^2+c_L m_Z^2 & m_t(A_t-\mu\cot\beta)\cr
 m_t(A_t-\mu\cot\beta)& M_{\widetilde U}^2+m_t^2+c_R m_Z^2\cr}\,,
\eqn\stopmatrix$$
where $c_L\equiv ({1\over 2}-{2\over 3}\sin^2\theta_W)\cos2\beta$,
$c_R\equiv {2\over 3}\sin^2\theta_W\cos2\beta$, $M_{\widetilde Q}$,
$M_{\widetilde U}$, and $A_t$ are soft-supersymmetry-breaking parameters,
and $\mu$ is the supersymmetric Higgs mass parameter.  Large mixing
requires that the off-diagonal terms above are of the same order as
the diagonal terms.  If there is large mixing in the third generation
squark sector, why not in the fourth generation squark sector as well?
In fact, if $A_{t^\prime}\simeq A_t$, the mixing in the fourth
generation squark sector would be too large, driving the smallest
eigenvalue of the ${\widetilde t}^\prime_L$--${\widetilde t}^\prime_R$
squared-mass matrix negative.  Remarkably, this does not occur
due to the infrared fixed-point behavior of the fourth generation.
$A_{t^\prime}$ is driven to a fixed point that is independent of
its high energy value.\refmark\poketal\
Roughly, $A_{t^\prime}\simeq -2m_{1/2}$
where $m_{1/2}$ is the high-energy (GUT-scale) value of the gaugino Majorana
mass.  In contrast, the top quark is not controlled by the infrared fixed point
(since in our model $m_t$ is not large enough); hence, $A_t$ can be chosen
large.
Moreover, choosing $\mu$ negative enhances the third generation
squark mixing while it somewhat suppresses the fourth generation
squark mixing.

4. If gaugino Majorana mass parameters are unified with a common GUT-scale
mass given by $m_{1/2}$,
then the gluino, chargino and neutralino masses are
determined by $m_{1/2}$, $\mu$, and $\tan\beta$.
Our model prefers the region of parameter space where
$m_{1/2}\ll|\mu|$ (with $\mu$ negative).
Then, our choice of $M_{\tilde\chi_1^0}\simeq 25$~GeV
fixes $m_{1/2}\simeq 55$~GeV.
Typical values for the masses of the other light chargino and
neutralino states are
$M_{\tilde\chi_1^\pm}\simeq M_{\tilde\chi_2^0}\simeq 60$~GeV.
The choice of $m_{1/2}$ also fixes the gluino mass;
we find $M_{\tilde g}\simeq 3m_{1/2}
\simeq 165$~GeV.  The dominant decay of this gluino would be
$\widetilde g\rta \widetilde t\bar t$ (or its charge-conjugated state).
Such a gluino cannot be ruled out by present Tevatron limits.

\REF\cleo{M.S. Alam \etal\ [CLEO Collaboration], {\sl Phys. Rev. Lett.}
{\bf 74} (1995) 2885.}
We have checked that virtual effects of
the light supersymmetric particles do not generate new conflicts
with experimental data.  For example, because the light chargino
is nearly a pure gaugino, the chargino--top squark loop has a
negligible effect on the rate for $Z\rta b\bar b$.  Our model then
predicts $R_b=0.2184$, which is within one standard deviation of the measured
LEP value [eq.~\zbbnumbers].  The improvement over the Standard Model
result is due to the fact that $m_t\simeq m_W$.
As a second example, one of the
most sensitive tests of the model is to check that its prediction for
$b\rta s\gamma$ is consistent with $1.0\times 10^{-4}\lsim
BR(b\rta s\gamma)\lsim 4\times 10^{-4}$, as required by the CLEO
measurement.\refmark\cleo\
The predictions of our model live comfortably within this bound.

\REF\radcorr{H.E. Haber and R. Hempfling, {\sl Phys. Rev. Lett.} {\bf 66}
(1991) 1815; {\sl Phys. Rev.} {\bf D48} (1993) 4280;
Y. Okada, M. Yamaguchi and T. Yanagida, {\sl Prog. Theor. Phys.}
{\bf 85} (1991) 1;  {\sl Phys. Lett.} {\bf B262} (1991) 54;
J. Ellis, G. Ridolfi and F. Zwirner, {\sl Phys. Lett.}
{\bf B257} (1991) 83; {\bf B262} (1991) 477; R. Barbieri, M. Frigeni,
and F. Caravaglios {\sl Phys. Lett.} {\bf B258} (1991) 167.}
5. The mass of the lightest CP-even Higgs boson should lie above the
LEP lower limit.  For $\tan\beta=1.6$, the tree-level {\it upper bound}
on the light Higgs mass is $\mhl\leq m_Z|\cos{2\beta}|=40$~GeV, which would
have been detected at LEP.  However, radiative corrections can raise the
upper bound substantially.\refmark\radcorr\
The bound increases with increasing
values of the soft-supersymmetry-breaking parameters which appear in
the squark squared-mass matrix [eq.~\stopmatrix].  We find as a typical
range of values that $\mhl\simeq 65$--70~GeV, above the present LEP limits.

6. The Tevatron may be able to rule out the existence of the $b^\prime$
with mass $m_{b^\prime}\simeq 110$~GeV.
If kinematically allowed, the decay $b^\prime\rta\widetilde
t\widetilde\chi_1^-$ would be the dominant decay mode.  If disallowed,
there would be a competition between $b^\prime\rta Wc$ (a change of two
generations) and $b^\prime\rta W^\star t$ (a change of one generation, but
suppressed by three-body phase space).  If necessary, one can choose
$|V_{b^\prime c}|\ll |V_{t^\prime b}|$ to remove the possibility of
$b^\prime\rta Wc$.  Then, all $b^\prime$ decays would result in
$W^\star c\widetilde\chi_1^0\widetilde\chi_1^0$.
There are no published limits that exclude such a
$b^\prime$.  However, a dedicated search at the Tevatron should be
able to discover or exclude such events.

7. Perhaps the most difficult requirement for our model is to
reproduce the oblique electroweak radiative corrections inferred from
the precision measurements at LEP.  Consider the contributions to
$\delta\rho$.  Since in our model,
$m_t$ is less than half of its standard value,
the contribution of the $t$--$b$ doublet to $\delta\rho$ is reduced
by a factor of 4.  This cannot be made up entirely by the contribution
of the fourth generation fermions, since the mass of the $b^\prime$ is
not negligible.  We find that the contributions of the third and
fourth generation fermions make up only half the observed $\delta\rho$.
The remainder must come from the third and fourth generation squarks.
This requirement places severe restrictions on the squark
parameters [eq.~\stopmatrix].
One must maximize the off-diagonal squark mixing while keeping
the diagonal squark mass parameters as small as possible.  However,
the latter cannot be too small; otherwise the radiative corrections to
the light Higgs mass will be reduced leading to a value of $\mhl$ below
the current LEP bound.

\REF\peskin{M.E. Peskin and T. Takeuchi, {\sl Phys. Rev. Lett.}
{\bf 65} (1990) 964; {\sl Phys. Rev.} {\bf D46} (1992) 381.}
\REF\langacker{J. Erler and P. Langacker, UPR-0632-T (1994)
[hep-ph 9411203]; P. Langacker, private communication.}
\REF\stable{J. Ellis, D.V. Nanopoulos and K. Tamvakis, {\sl Phys. Lett.}
{\bf 121B} (1983) 123; L. Iba\~nez and C. Lopez, {\sl Phys. Lett.}
{\bf 126B} (1983) 54; L. Alvarez-Gaum\'e, J. Polchinski and M. Wise,
{\sl Nucl. Phys.} {\bf B221} (1983) 495.}
It is convenient to parameterize the oblique radiative corrections in terms
of the Peskin-Takeuchi variables\refmark\peskin\
$S$, $T$ and $U$.  Here $T\equiv \alpha^{-1}
\delta\rho$ (where $\alpha^{-1}\simeq 137$) is the the most sensitive
(although some interesting restrictions can be obtained by considering $S$).
Langacker has performed a global analysis of precision electroweak
data,\refmark\langacker\
assuming that $m_t=80$~GeV and $\mhl=65$~GeV, and extracts values
for the oblique parameters.  He finds $T_{\rm new}= 0.70\pm 0.21$, which
in our model must arise from the contribution of the fourth
generation fermions and the third and fourth generation squarks.
(The contributions from other supersymmetric particles are negligible.)
We find that the fourth generation fermions yield a contribution of 0.2
to $T_{\rm new}$.  The contributions of the third and fourth generation
squarks depend sensitively on the squark parameters as noted above;
a range of parameters can be found that yields
a total squark contribution to $T_{\rm new}$
that lies between 0.3 and 0.4.  This would bring us within
one standard deviation of Langacker's value for $T_{\rm new}$.
To achieve such a value for the squark contribution to
$T_{\rm new}$ requires substantial
$\widetilde q_L$--$\widetilde q_R$ mixing in the third
generation, which is uncomfortably large and may cause
stability problems\refmark\stable\
for the complete scalar potential of the model.
Non-negligible mixing in the fourth generation also enhances the
fourth generation squark contributions to $T_{\rm new}$.  The maximum
effect is limited phenomenologically by
a lower bound on the mass of $\widetilde b^\prime$.
In order that $t^\prime\rta bW^+$ remain the dominant decay, one
must kinematically forbid $t^\prime\rta\widetilde b^\prime\widetilde
\chi_1^+$.  Given $M_{\widetilde\chi_1^\pm}\simeq 60$~GeV,
a value of $M_{\tilde b^\prime}\simeq 120$~GeV
is a comfortable choice.  All the phenomenological
constraints have now forced
the parameters of the model into a very narrow corner of parameter space.

\chapter{Conclusions}
\medskip

It is still possible that $m_t\simeq m_W$, despite the recent announcement
of the top quark discovery by the CDF and D0 collaborations.   A
model has been exhibited that satisfies all phenomenological constraints
and is not ruled out by published data.  The most
theoretically troubling feature of the model is the large mixing among
the third generation squarks that is necessary to ensure a viable
prediction for the electroweak $\rho$-parameter.

The model possesses a rich spectrum of new particles that will be accessible
to LEP-II and the Tevatron.   In particular, eight new particles of this
model could be discovered at LEP-II: the $t$-quark, the fourth generation
leptons ($\tau^\prime$ and $N$), the light Higgs
boson ($\hl$), and four supersymmetric particles ($\widetilde\chi_1^0$,
$\widetilde\chi_2^0$, $\widetilde\chi_1^\pm$, and $\widetilde t$).
Note that even at the initial run of LEP-II at $\sqrt{s}=140$~GeV planned
for the fall of 1995, all four supersymmetric particles listed above
(and the $\tau^\prime$) should be discovered, or else the model would be
excluded.

\REF\pois{J.F. Gunion, D.W. McKay and H. Pois, {\sl Phys. Lett.}
{\bf B334} (1994) 339.}
Thus, the fate of this model may be decided before these Proceedings
appear in print.  Nevertheless, this exercise was useful in
demonstrating the difficult in constructing four-generation models
of low-energy supersymmetry.
In a previous work, Gunion, McKay and Pois\refmark\pois\
attempted to construct four-generation models in the context of minimal
low-energy supergravity.  They identified the top quark as the
state discovered by CDF and D0.  In order to keep Higgs-quark
Yukawa couplings perturbative up to the GUT scale, they
were forced to try to hide the $b^\prime$ and $t^\prime$ in a mass region
below $m_t\simeq 175$~GeV.  The resulting models were contrived and
phenomenologically unappealing.  Our approach represents the logical
alternative for four-generation low-energy supersymmetric models.
If these models are excluded, one will finally be able to state with
confidence that in the low-energy suersymmetric approach the number
of generations is indeed three!

\endpage
\centerline{\bf Acknowledgments}
\medskip
I would like to
thank Marcela Carena and Carlos Wagner for an enjoyable and
rewarding collaboration.  I am also grateful to Jean-Marie Fr\`ere
for his kind invitation to speak at the Moriond meeting.
Finally, I send
a special appreciation to Jo\"elle Raguideau, whose encouragements
were a great help to a painful knee.  This work was supported in
part by a grant from the U.S. Department of Energy.
\bigskip
\refout

\bye